\documentclass[amsmath, amssymb, showpacs, reprint, floatfix, aps, prl, superscriptaddress]{revtex4-1}
\usepackage{graphicx}
\usepackage{dcolumn}
\usepackage{physics}
\usepackage{braket}
\usepackage{bm}
\usepackage{color}
\usepackage{hyperref}
\hypersetup{colorlinks=true,citecolor=blue,linkcolor=blue, urlcolor=blue}
\newcommand{\kk}{\mathbf{k}}
\renewcommand{\qq}{\mathbf{q}}

\begin{document}
\title{Piezoelectric Electron-Phonon Interaction from \textit{Ab Initio} Dynamical Quadrupoles: Impact on Charge Transport in Wurtzite GaN}
\author{Vatsal A. Jhalani}
\author{Jin-Jian Zhou}
\author{Jinsoo Park}
\affiliation{Department of Applied Physics and Materials Science, Steele Laboratory, California Institute of Technology, Pasadena, California 91125, USA.}
\author{Cyrus E. Dreyer}
\affiliation{Department of Physics and Astronomy, Stony Brook University, Stony Brook, New York 11794-3800}
\affiliation{Center for Computational Quantum Physics, Flatiron Institute, 162 Fifth Avenue, New York, New York 10010}
\author{Marco Bernardi}
\email[Corresponding author: ]{bmarco@caltech.edu}
\affiliation{Department of Applied Physics and Materials Science, Steele Laboratory, California Institute of Technology, Pasadena, California 91125, USA.}
%
%
\begin{abstract}
First-principles calculations of $e$-ph interactions are becoming a pillar of electronic structure \mbox{theory.} However, the current approach is incomplete. 
The piezoelectric (PE) $e$-ph interaction, a long-range scattering mechanism due to acoustic phonons in non-centrosymmetric polar materials, %
is not accurately described at present. Current calculations include short-range $e$-ph interactions (obtained by interpolation) and the dipole-like Fr\"ohlich long-range coupling in polar materials, 
but lack important quadrupole effects for acoustic modes and PE materials.    
Here we derive and \mbox{compute} the long-range $e$-ph interaction due to dynamical quadrupoles, and apply this framework to investigate $e$-ph interactions and the carrier mobility in the PE material wurtzite GaN.
We show that the quadrupole contribution is essential to obtain accurate $e$-ph matrix elements for acoustic modes and to compute PE scattering. 
Our work resolves the outstanding problem of correctly computing $e$-ph interactions for acoustic modes from first principles, and enables studies of $e$-ph coupling and charge transport in PE materials. 
\end{abstract}
\maketitle
%
%
When atoms move due to lattice vibrations, the potential seen by an electron quasiparticle changes due to both short-range and long-range forces. 
Early theories of such electron-phonon ($e$-ph) interactions focused on simplified models tailored to specific materials~\cite{Ziman}. 
In metals, the so-called \lq\lq deformation potential'' quantifies the short-range $e$-ph interactions with acoustic phonons~\cite{Bardeen}. 
In ionic and polar covalent materials, in which atoms can be thought of as carrying a net charge, Fr\"ohlich identified a dipole-like long-range $e$-ph interaction due to longitudinal optical (LO) phonons~\cite{Frohlich}. %
In polar materials lacking inversion symmetry, the piezoelectric (PE) $e$-ph interaction, due to strain induced by acoustic phonons, is also important~\cite{Mahan, Mahan-2}.  
Its conventional formulation by Mahan expresses the $e$-ph coupling in terms of the macroscopic PE constants of the material~\cite{Mahan-2}.
\\ 
\indent
Vogl unified these $e$-ph interactions~\cite{Vogl1976}, showing that the dipole, PE, and deformation-potential contributions originate from a multipole expansion~\cite{Lawaetz} of the $e$-ph potential,  
and analyzed its behavior in the long-wavelength limit (phonon wave vector $\qq \!\rightarrow\! 0$). The Fr\"ohlich dipole term diverges as $1/q$ in this limit, %
and is proportional to the sum of atomic dipoles: If each atom $\kappa$ is associated with a Born charge tensor, $\mathbf{Z}_{\kappa}$, 
contracting this tensor with the phonon eigenvector $\mathbf{e}_{\nu \qq}$ gives the atomic contributions to the dipole field, $\mathbf{Z}_{\kappa}\, \mathbf{e}_{\nu \qq}$. 
Next in the multipole expansion is the quadrupole field generated by the atomic motions, which approaches a constant value as $\qq \rightarrow 0$. %
If each atom is associated with a dynamical quadrupole tensor, $\mathbf{Q}_{\kappa}$, the atomic quadrupole contributions can be written as  
$\mathbf{Q}_{\kappa} \mathbf{e}_{\nu \qq}$. Both the dipole and the quadrupole terms contribute to the PE $e$-ph interaction~\cite{Vogl1976}.
The multipole expansion also gives octopole and higher terms, which vanish at $\qq \!\rightarrow\! 0$ and can be grouped together into a short-range deformation potential. 
\\
\indent
Density functional theory (DFT)~\cite{Martin-book} and density functional perturbation theory (DFPT)~\cite{Baroni} have enabled calculations of $e$-ph interactions from first principles.  
In turn, the $e$-ph interactions can be used in the Boltzmann transport equation (BTE) framework to predict electron scattering processes and charge transport~\cite{Bernardi2014, Bernardi-noble, Zhou2016, Jhalani2017, Lee2018, Zhou-STO, Zhou-polaron, Perturbo, Li2015, Chen2017, Li2018, Sohier2018}. %
The recent development of the \textit{ab initio} Fr\"ohlich $e$-ph interaction~\cite{Sjakste2015, Verdi-frohlich} 
has been a first step toward implementing Vogl's modern $e$-ph theory in first-principles calculations and correctly capturing the long-range $e$-ph contributions.  
However, a key piece is still missing in the \textit{ab initio} framework: the quadrupole $e$-ph interaction, which critically corrects the dipole term in polar materials, 
is sizable in nonpolar materials, and is particularly important in PE materials such as wurtzite crystals or titanates. 
\\
\indent 
In this Letter, we derive the \textit{ab initio} quadrupole $e$-ph interaction and compute it for a PE material, wurtzite gallium nitride (GaN), using dynamical quadrupoles computed from first principles.  
We show that including the quadrupole term provides accurate $e$-ph matrix elements and is essential to obtaining the correct acoustic phonon contribution to carrier scattering. 
We compute the electron and hole mobilities in GaN including the quadrupole interaction, obtaining results in agreement with experiment.  
Our analysis highlights the large errors resulting from including only the Fr\"ohlich term in GaN~\cite{PonceGaNPRB}, which greatly overestimates the acoustic mode $e$-ph interactions~\cite{PonceGaNPRB,PonceGaNPRL}. 
Our companion paper applies this framework to silicon and the PE material PbTiO$_3$. The $e$-ph quadrupole contribution is sizable in both cases, and essential to correctly compute the $e$-ph matrix elements. 
Taken together, the quadrupole interaction completes the theory and enables accurate \textit{ab initio} $e$-ph calculations in all materials and for all phonon modes.
\\
\indent 
The key ingredient in first-principles $e$-ph calculations are the $e$-ph matrix elements, $g_{mn\nu}(\kk,\qq)$, which encode the probability amplitude for scattering from an initial Bloch state $\ket{n\kk}$ with band index $n$ and crystal momentum $\kk$ into a final state $\ket{m\kk+\qq}$ by emitting or absorbing a phonon with branch index $\nu$ and wave vector $\qq$~\cite{Bernardi-review, Perturbo}. 
Following Vogl~\cite{Vogl1976}, we separate the long-range dipole and quadrupole contributions from the short-range part:
\begin{equation}\label{eq:gtot}
g_{mn\nu}(\kk,\qq) = g^\text{dipole}_{mn\nu}(\kk,\qq) + g^\text{quad}_{mn\nu}(\kk,\qq) + g^\text{S}_{mn\nu}(\kk,\qq)
\end{equation}
where $g^\text{dipole}$ is the \textit{ab initio} Fr\"ohlich interaction written in terms of Born effective charges~\cite{Sjakste2015, Verdi-frohlich}, 
$g^\text{quad}$ is the quadrupole interaction written in terms of dynamical quadrupoles, and 
$g^\text{S}$ collects octopole and higher-order short-range terms. 
For polar acoustic modes, $g^\text{dipole} + g^\text{quad}$ is a generalized replacement for the phenomenological PE $e$-ph coupling expressed in terms of the PE constants~\cite{Mahan, Mahan-2}.  
We derive the first-principles quadrupole $e$-ph interaction $g^\text{quad}_{mn\nu}(\kk,\qq)$ by superimposing two oppositely oriented dipole moments, as discussed in detail in our companion paper, and obtain: 
\begin{widetext}
\begin{equation}\label{eq:quad}
g_{mn\nu}^\text{quad}(\kk, \qq) = \frac{e^2}{\Omega\,\varepsilon_{0}} \sum_{\kappa} \left(\frac{\hbar}{2\omega_{\nu \qq} M_\kappa}\right)^{\frac{1}{2}} \sum_{\mathbf{G}\neq-\qq}  \frac{ \frac{1}{2} (\mathrm{q}_\alpha+\mathrm{G}_\alpha) (Q_{\kappa,\beta}^{\alpha\gamma} \mathbf{e}_{\nu \qq, \beta}^{(\kappa)} ) (\mathrm{q}_\gamma+\mathrm{G}_\gamma) }{(\mathrm{q}_\alpha+\mathrm{G}_\alpha)\epsilon_{\alpha\gamma} (\mathrm{q}_\gamma+\mathrm{G}_\gamma)} \bra{m\kk+\qq} 
e^{i(\mathbf{q}+\mathbf{G})\cdot(\mathbf{r}-\mathbf{\tau}_{\kappa})} \ket{n\kk},
\end{equation}
\end{widetext}
where $e$ is the electron charge, $\Omega$ is the unit cell volume, $M_\kappa$ and $\bm{\tau}_\kappa$ are the mass and position of the atom with index $\kappa$, 
$\mathbf{G}$ are reciprocal lattice vectors, $\mathbf{e}_{\nu \qq}^{(\kappa)}$ is the phonon eigenvector projected on atom $\kappa$, and $\boldsymbol{\epsilon}$ is the 
dielectric tensor of the material. 
Summation over the Cartesian indices $\alpha$, $\beta$, $\gamma$ is implied. 
\\
\indent
The dynamical quadrupoles $\bm{Q}_{\kappa}$ entering Eq.~(\ref{eq:quad}) are third-rank tensors defined as the second order term in the long-wavelength expansion of the cell-integrated charge-density response to a monochromatic displacement~\cite{Martin1972, Stengel2013, Royo2019}. 
Here, they are computed by symmetrizing the first-order-in-\textbf{q} polarization response~\cite{Stengel2013, Dreyer2018}:
\begin{equation}\label{eq:Q2}
  Q^{\alpha\gamma}_{\kappa,\beta}=i\Omega\left(\frac{\partial \overline{P}^{\textbf{q}}_{\alpha,\kappa\beta}}{\partial q_\gamma}\Bigg\vert_{\textbf{q}=0}+\frac{\partial \overline{P}^{\textbf{q}}_{\gamma,\kappa\beta}}{\partial q_\alpha}\Bigg\vert_{\textbf{q}=0}\right)
\end{equation}
\\
\indent
With the full first-principles $e$-ph matrix elements in Eq.~(\ref{eq:gtot}) in hand, we compute the phonon-limited mobility at temperature $T$ using 
the BTE in both the relaxation time approximation (RTA) and with a full iterative solution~\cite{Perturbo}.  
Briefly, we compute the $e$-ph scattering rates (and their inverse, the relaxation times $\tau_{n\kk}$), from the lowest order $e$-ph self-energy~\cite{Bernardi-review}.
The mobility is then obtained as the energy integral~\cite{Perturbo}:
\begin{equation} \label{eq:mob}
\mu_{\alpha\beta}(T) = \frac{e}{n_c\Omega N_{\kk}} \int\!\mathrm{d}E\left(\!-\frac{\partial f}{\partial E}\right)\sum_{n\kk} \bm{F}^\alpha_{n\kk}(T) \mathbf{v}^\beta_{n\kk}\,\delta(E - \varepsilon_{n\kk})
\end{equation}
where $n_c$ is the carrier concentration, $f$ is the Fermi-Dirac distribution, $N_\kk$ is the number of $\kk$-points, 
and $\mathbf{v}_{n\kk}$ and $\varepsilon_{n\kk}$ are electron band velocities and energies, respectively.    
The $\bm{F}^\alpha_{n\kk}$ term is obtained as $\tau_{n\kk}(T) \mathbf{v}^\alpha_{n\kk}$ in the RTA or by solving the BTE iteratively.
\\
\indent  
%
%
We carry out \textit{ab initio} calculations on wurtzite GaN with relaxed lattice parameters, using the same settings as in our recent work~\cite{GaNLifetime}. 
The ground state properties and electronic wave functions are computed using DFT in the generalized gradient approximation~\cite{GGA} with the \textsc{Quantum ESPRESSO} code~\cite{QE-2009, QE-2017}. 
We include spin-orbit coupling by employing fully relativistic norm-conserving pseudopotentials~\cite{ONCVPSP,PBEsol} (generated with Pseudo Dojo \cite{Dojo}) and correct the DFT band structure using $GW$ results. 
We use DFPT~\cite{Baroni} to compute phonon frequencies and eigenvectors, and obtain the $e$-ph matrix elements $g_{nm\nu}(\kk,\qq)$ on coarse $8 \times 8 \times 8$ $\kk$-point and $\qq$-point grids~\cite{mp1976} using our {\sc{Perturbo}} code~\cite{Perturbo}. 
We obtain the dynamical quadrupoles $\mathbf{Q}_{\kappa}$ by computing $\overline{P}^{\textbf{q}}_{\alpha,\kappa\beta}$ in Eq.~(\ref{eq:Q2}) with the methodology in Ref.~[\onlinecite{Dreyer2018}] as implemented in the \textsc{ABINIT} code~\cite{ABINIT}, and validate the results against DFPT clamped-ion PE constants (see Supplemental Material~\cite{supp-info}). 
We subtract the long-range terms $g^\text{dipole} + g^\text{quad}$, interpolate the $e$-ph matrix elements using Wannier functions~\cite{Wann90} to fine $\kk$- and $\qq$-point grids, and then add back the long-range terms. 
The resulting matrix elements are employed to compute relaxation times and mobilities~\cite{supp-info} with the {\sc{Perturbo}} code~\cite{Perturbo}.
\\
\indent 
%
%
%
%
\begin{figure}
\includegraphics[width=0.9\linewidth]{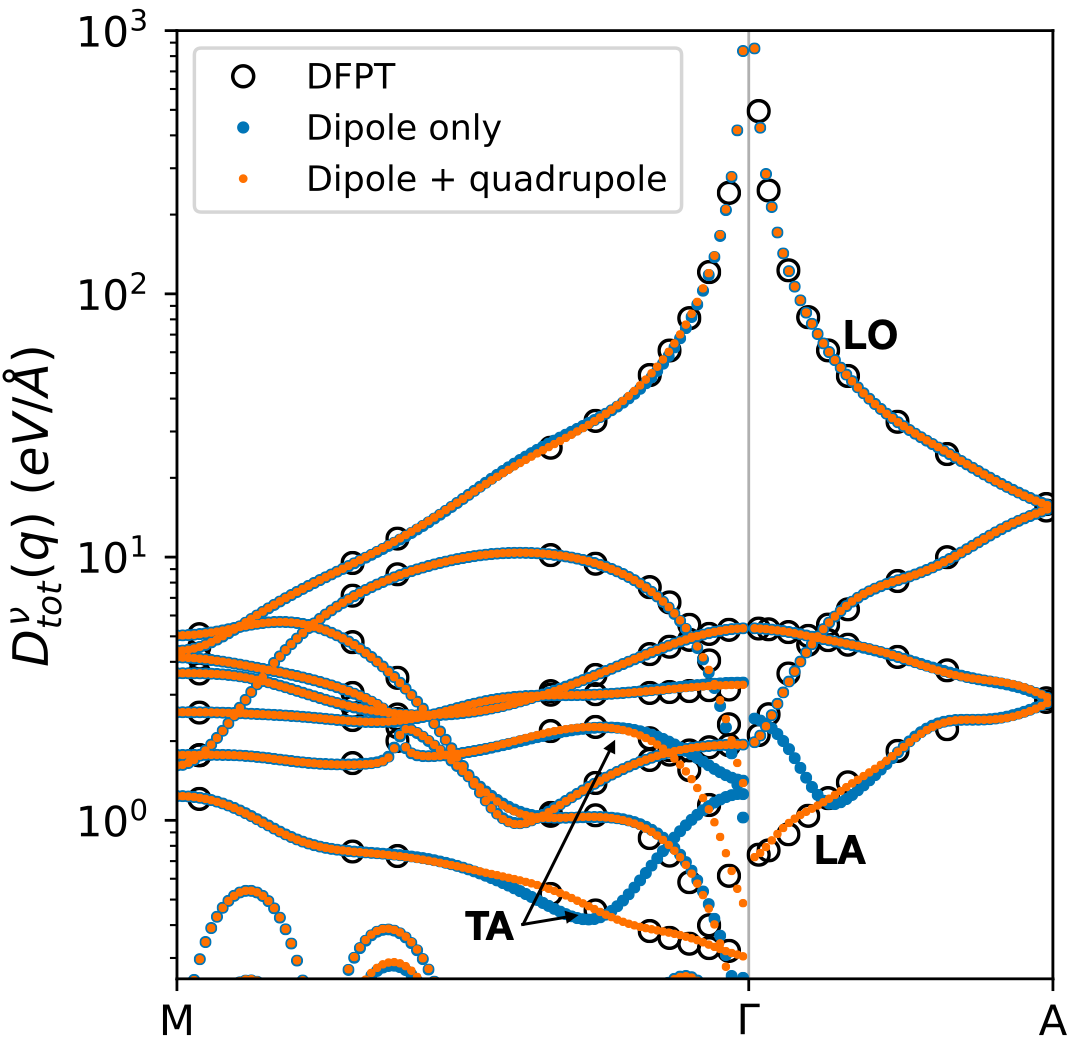}
\caption{  
Mode-resolved $e$-ph coupling strength, $D^\nu_\text{tot}(\qq)$, computed using the two lowest conduction bands; the initial state is fixed at $\Gamma$ and $\qq$ is varied along high-symmetry lines. 
The $D^\nu_\text{tot}(\qq)$ data computed with $e$-ph matrix elements from DFPT (black circles) is used as a benchmark, 
and compared with Wannier interpolation plus Fr{\"o}hlich interaction (blue) or plus Fr{\"o}hlich and quadrupole interactions (orange).
}\label{quadcompare}
\end{figure}
Since DFPT can accurately capture the long-range $e$-ph interactions, it can be used as a benchmark for our approach of including the dipole plus quadrupole terms after interpolation. 
Note that due to computational cost, DFPT calculations cannot be carried out on the fine grids needed to compute electrical transport, 
so the long-range terms need to be added after interpolation.  
Following Ref.~\cite{Sjakste2015}, we define a gauge-invariant $e$-ph coupling strength, $D^\nu_\text{tot}$, proportional to the absolute value of the $e$-ph matrix elements:
\begin{equation}
D^{\nu}_{\mathrm{tot}}\left(\mathbf{q}\right) = \hbar^{-1}\sqrt{ 2 \omega_{\nu\mathbf{q}}M_{\mathrm{uc}} \sum_{mn}\left|g_{mn\nu}\left(\mathbf{k}=\Gamma,\mathbf{q}\right)\right|^{2}/N_{b}}\,,
\label{eq:g_abs}
\end{equation}
with $M_\text{uc}$ the unit cell mass and $N_b$ the number of bands. We compute $D^{\nu}_{\mathrm{tot}}\left(\mathbf{q}\right)$ with various approximations to analyze the role of the quadrupole $e$-ph interactions.
\\
\indent
In Fig.~\ref{quadcompare}, we use $D^\nu_\text{tot}(\qq)$ obtained from direct DFPT calculations of the matrix elements as a benchmark, 
and compare calculations that include only the long-range Fr\"ohlich dipole interaction and both the dipole and quadrupole interactions. 
The short-range $e$-ph interactions are included in both cases as a result of the Wannier interpolation. 
Including the quadrupole term dramatically improves the accuracy of the $e$-ph matrix elements for the longitudinal acoustic (LA) and transverse acoustic (TA) modes at small $\qq$ (near $\Gamma$ in Fig.~\ref{quadcompare}).
The discrepancy between the dipole-only calculation and DFPT is completely corrected when including the quadrupole term, which reproduces the DFPT benchmark exactly. 
While the dipole-only scheme leads to large errors, the dipole and quadrupole contributions, as can be seen, cancel each other out since they are nearly equal and opposite for acoustic modes in GaN. 
\\
\indent
Both the dipole and quadrupole terms contribute to the PE $e$-ph interaction from the LA and TA acoustic modes~\cite{Vogl1976}. Expanding the phonon eigenvectors at $\qq \!\rightarrow\! 0$ 
as $\mathbf{e}_{\nu\qq} \!\approx\! \mathbf{e}^{(0)}_{\nu\qq} + i q\, \mathbf{e}^{(1)}_{\nu\qq}$~\cite{Vogl1976}, one finds two PE contributions of order $q^0$ for $\qq \!\rightarrow\! 0$~\cite{Vogl1976}. %
One is from the Born charges, $\mathbf{Z}_{\kappa} (i q\, \mathbf{e}^{(1)}_{\nu \qq})$, and is a dipole-like interaction generated by atoms with a net charge experiencing different displacements due to an acoustic mode. 
The other is from the dynamical quadrupoles, $\mathbf{Q}_{\kappa} \mathbf{e}^{(0)}_{\nu\qq}$, and is associated with a clamped-ion electronic polarization~\cite{Bernardini1997}. %
As a result, the \textit{ab initio} Fr\"ohlich term includes only part of the PE $e$-ph interaction, so the dipole-only scheme fails in GaN because it neglects the all-important quadrupole electronic contribution.  
%
%
We also implemented and tested Mahan's phenomenological PE coupling~\cite{Mahan-2}, 
\begin{equation}\label{eq:piezo}
\tilde{g}^{\rm PE}_\nu(\qq) = 4\pi\frac{e^2}{4\pi\varepsilon_0}
\left[ \frac{\hbar}{2\omega_{\nu\qq}M_{\rm uc}}\right]^\frac{1}{2}
\frac{q_\alpha e_{\alpha,\beta\gamma}  \mathbf{e}_{\nu \qq, \beta} q_\gamma}{q_\alpha \epsilon_{\alpha\gamma} q_\gamma}
\end{equation}
where $e_{\alpha,\beta\gamma}$ are the PE constants of GaN~\cite{supp-info}. In the $\qq \!\rightarrow\! 0$ limit, this model includes both ionic-motion and electronic effects~\cite{Vogl1976}, 
and is a less computationally demanding alternative that does not require computing the dynamical quadrupoles. 
We find that the $e$-ph coupling $D^\nu_\text{tot}(\qq)$ obtained from the Mahan model improves over the dipole-only scheme~\cite{supp-info}, but exhibits discrepancies with direct DFPT calculations  
at finite $\qq$, and overall is inadequate for quantitative calculations. 
\\
\indent
%
%
\begin{figure}[b]
\includegraphics[width=1.0\linewidth]{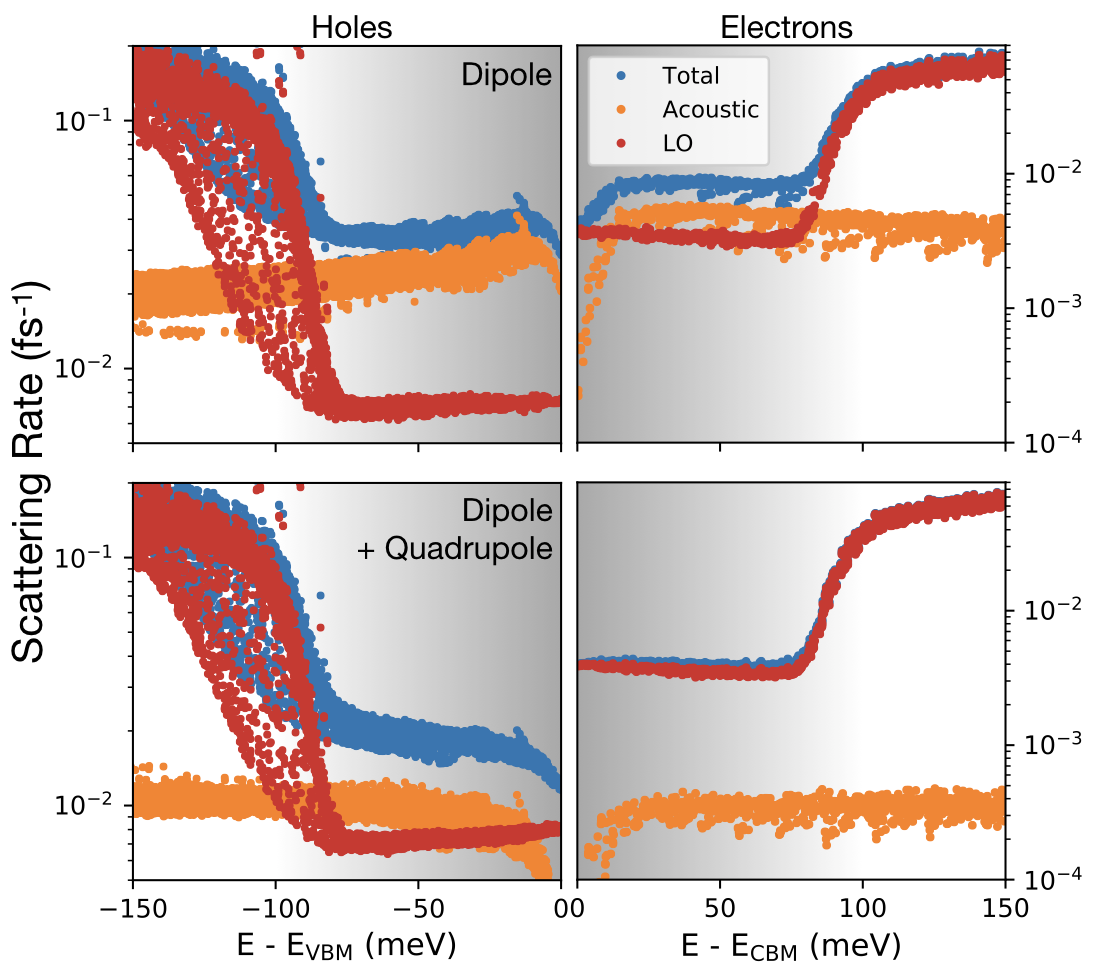}
\centering
\caption{
Electron-phonon scattering rates at 300~K. 
We compare calculations including the long-range Fr{\"o}hlich interaction only (top) with results including the Fr{\"o}hlich and quadrupole interactions (bottom).
The short-range $e$-ph interactions are included in both cases as a result of the interpolation. We plot the total scattering rate (blue) as well as the contributions from the LO (red) and acoustic (orange) modes for holes (left) and electrons (right) as a function of carrier energy within 150 meV of the band edges. The gray shading represents the energies at which carriers contribute to the mobility, as given by the integrand in Eq.~(\ref{eq:mob}). 
The hole and electron energy zeros are the valence and conduction band edges, respectively.
}\label{scatrates}
\end{figure}
Figure~\ref{scatrates} shows the effect of using the more accurate quadrupole scheme on the $e$-ph scattering rates (defined as the inverse of the $e$-ph relaxation times, $\Gamma_{n\kk} = 1/ \tau_{n\kk}$) computed at 300 K.
%
%
We focus on the energy range of interest for charge transport near 300 K, namely an energy window within $\sim$100 meV of the band edges. 
Since these energies are below the LO phonon emission threshold (90 meV in GaN), LO scattering is suppressed and dominated by thermally activated LO phonon absorption processes. 
On the other hand, there is a large phase space for acoustic phonon scattering, especially for interband processes in the valence band. 
For electrons, including the quadrupole term greatly suppresses the acoustic mode contribution to the scattering rates, reducing it from nearly half of the total scattering rate to a negligible contribution (Fig.~\ref{scatrates}). 
This result is due to the cancellation of the dipole and quadrupole terms for acoustic phonons discussed above.  %
As expected, the LO contribution becomes dominant for electrons in the conduction band, where small-$\mathbf{q}$ intravalley scattering is controlled by the Fr\"ohlich interaction with 1/q behavior rather than 
by the PE dipole and quadrupole terms, both of which have a $q^0$ trend at small $\mathbf{q}$. Using correct $e$-ph matrix elements that include the quadrupole term, our calculation restores this physical intuition.\\  
%
%
\begin{figure}[!t]
\includegraphics[width=0.9\linewidth]{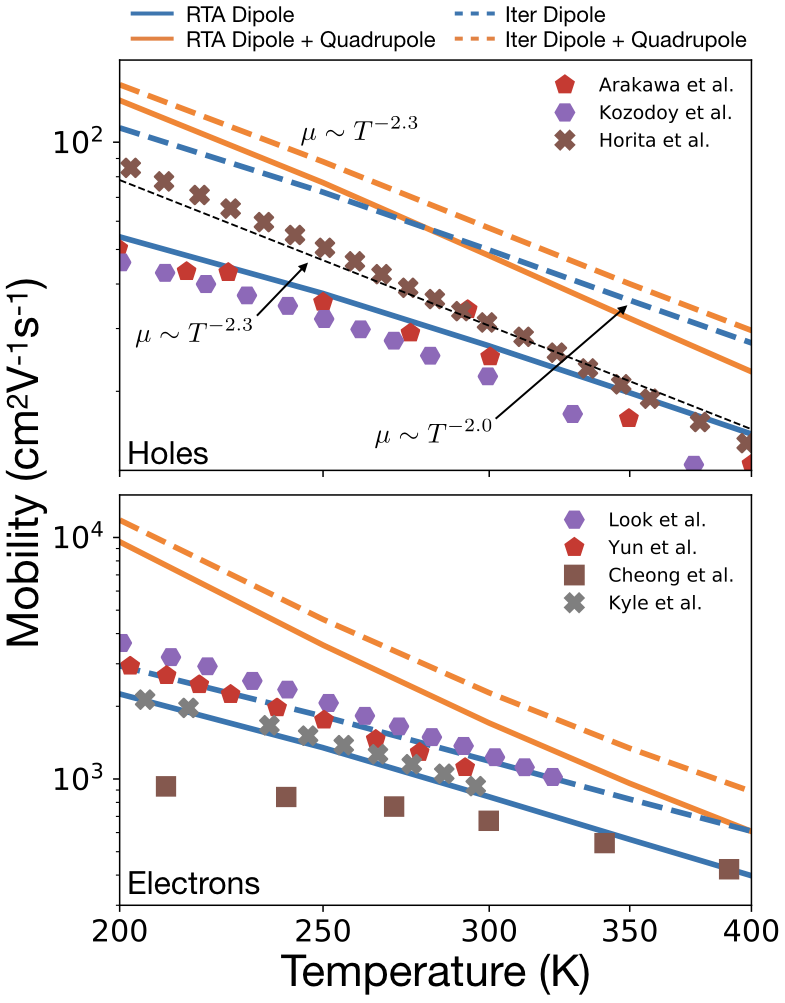}
\caption{
Electron (top) and hole (bottom) mobilities in wurtzite GaN, computed in the [1000] plane. We show our computed results obtained with the RTA (solid lines) and iterative BTE (dashed lines), 
both with the long-range Fr\"ohlich interaction only (blue) and with dipole plus quadrupole interactions (orange). Experimental results from Refs.~\cite{Look2001, Yun2000, Cheong2000, Kyle2014} for electrons and Refs.~\cite{Arakawa2016, Kozodoy2000, Horita2017} for holes are shown for comparison.
}\label{mobilities}
\end{figure}
\indent
A similar but less pronounced trend is found for holes in Fig.~\ref{scatrates}, where at the peak of the mobility integrand ($\sim$50 meV below the valence band edge) acoustic phonon scattering is suppressed
from 75\% of the total scattering rate when only the Fr\"ohlich interaction is included to less than 50\% when including the quadrupole term. 
The large acoustic phonon scattering for holes found in recent calculations~\cite{PonceGaNPRB} is partially due to the fact that the quadrupole term was missing. 
The possibility of increasing hole mobility by engineering lower acoustic scattering based on these investigations~\cite{PonceGaNPRL} clearly does not properly take into account PE scattering 
(and PE charges induced by straining GaN), and should be revisited.
\\
\indent 
%
%
Analysis of the temperature dependent mobility in GaN, computed using Eq.~(\ref{eq:mob}) with both the RTA and iterative BTE approaches, 
highlights the key role of the quadrupole $e$-ph interaction. %
Figure~\ref{mobilities} shows the electron and hole mobilities in the basal [1000] plane of GaN, computed both using Wannier interpolation plus the Fr{\"o}hlich interaction 
and with our improved scheme including the quadrupole term. 
Experimental mobility measurements~\cite{Look2001, Yun2000, Cheong2000, Kyle2014, Arakawa2016, Kozodoy2000, Horita2017} are also given for comparison. %
Compared to calculations that include only the dipole Fr\"ohlich interaction, including the quadrupole term removes the artificial overestimation of acoustic phonon scattering, 
thus increasing the computed mobility and correcting its temperature dependence, especially at lower temperatures, where acoustic scattering is dominant.
\\
\indent
We find good agreement between our computed electron and hole mobilities and experimental results, especially when comparing with  
the highest mobilities measured in samples with low doping concentrations ($\sim$$10^{15}\text{ cm}^{-3}$ for n-type \cite{Look2001} and $\sim$$10^{16}-10^{17} \text{ cm}^{-3}$ for p-type \cite{Suda2016, Arakawa2016} GaN). 
In these high purity samples, charge transport is governed by phonon scattering in the temperature range we investigate, 
so these measurements are ideal for comparing with our phonon-limited mobilities.\\
\indent
We focus on the iterative BTE results, whose accuracy is superior to the RTA, also given for completeness~\footnote{For polar materials, the RTA mobility is typically lower than the more accurate mobility obtained with the iterative BTE, consistent with our results.}. 
For holes, in which acoustic scattering is significant, the temperature dependence of the mobility is improved after including the quadrupole interaction, 
as shown in the upper panel of Fig.~\ref{mobilities}. The exponent $n$ in the temperature dependence of the mobility, $\mu \sim T^{-n}$, is $n=2.3$ after including the quadrupole term,  
the same value as the exponent obtained by fitting the experimental data (for comparison, $n=2.0$ in the dipole-only calculation). %
The temperature dependence of the electron mobility is only in reasonable agreement with experiment. 
Including two-phonon processes may be needed to correctly predict the temperature trend, as we found recently for electrons in GaAs~\cite{Lee2019}, which similar to GaN has dominant LO phonon scattering in the conduction band.
\\ 
\indent
Our improved scheme increases the computed mobilities, placing them slightly above the experimental values. 
This trends is physically correct $-$ our computed mobilities are an upper bound %
as experimental samples may exhibit additional scattering from defects and interfaces. In addition, %
a slight mobility overestimation is expected in polar materials because including two-phonon scattering processes would lower the mobility and bring it closer to experiment, as we have recently shown in GaAs~\cite{Lee2019}.  
\\
\indent
In summary, we have presented a framework for computing the quadrupole $e$-ph interaction and including it in \textit{ab initio} $e$-ph calculations. Our results show its crucial contribution to acoustic phonon and PE scattering. 
Our work enables accurate calculations of long-range acoustic phonon interactions and paves the way to studies of charge transport in PE materials.
\vspace{5pt}
\begin{acknowledgments}
\noindent
V.J. thanks the Resnick Sustainability Institute at Caltech for fellowship support. J.P. acknowledges support by the Korea Foundation for Advanced Studies. 
This work was supported by the National Science Foundation under Grants No. CAREER-1750613 for theory development and ACI-1642443 for code development. 
J.-J.Z. acknowledges partial support from the Joint Center for Artificial Photosynthesis, a DOE Energy Innovation Hub, as follows: the development of some computational methods employed in this work was supported through the Office of Science of the U.S. Department of Energy under Award No. DE-SC0004993. C.E.D. acknowledges support from the National Science Foundation under Grant No. DMR-1918455. The Flatiron Institute is a division of the Simons Foundation. 
This research used resources of the National Energy Research Scientific Computing Center, a DOE Office of Science User Facility supported by the Office of Science of the U.S. Department of Energy under Contract No. DE-AC02-05CH11231.\\
\end{acknowledgments}

\textit{Note added.}$-$ While writing this manuscript, we became aware of a related preprint by another group~\cite{Brunin2020}. 
Their article analyzes how the quadrupole term improves $e$-ph matrix element interpolation, while ours focuses more broadly on the physics of $e$-ph interactions, acoustic phonons, and piezoelectric materials.

%

\end{document}